\documentstyle[12pt,psfig]{l-aa}
\input tcilatex
\newcommand{\beq}{\begin{equation}}
\newcommand{\eeq}{\end{equation}}
\newcommand{\ba}{\begin{array}}
\newcommand{\ea}{\end{array}}

\begin{document}
\pretolerance=5000

\title{Angular size in ``quintessence" cosmology}
\author{J. A. S. Lima and J. S. Alcaniz}

\institute{Departamento de Fisica, UFRN, C.P 1641\\
	   59072-970 Natal, Brasil }

\date{Received ; accepted}    

\offprints{limajas@dfte.ufrn.br}      

\maketitle

\begin{abstract} 
We investigate the influence of an exotic fluid component (``quintessence") on the angular 
size-redshift relation for distant extragalactic sources. Particular emphasis
is given for the redshif $z_{m}$ at which the angular size takes its minimal
value. We derive an analytical closed form which determines how $z_m$ depends
on the parameter of the equation of state describing the exotic component. The results for a flat 
model dominated by a ``quintessence"  are compared in
detail with the ones for the standard open model dominated by cold dark matter. Some 
consequences of systematic evolutionary effects on the values of $z_{m}$ are
also briefly discussed. It is argued that the critical redshift, for all
practical purposes, may completely be removed if such effects are taken into
account.   
\end{abstract}

\section{Introduction}

Recent data from SNe Ia have provided strong 
evidence for an expanding Universe speeding up, rather than 
slowing down (Riess et al. 1998; Perlmutter et al. 1998).
These observational evidences have stimulated great interest in 
a more general class of cosmological models driven by 
nonrelativistic 
matter and a ``quintessence" component, i.e., an exotic fluid 
with an arbitrary equation of 
state $p_{x} = \omega_x \rho_x$ ($\omega_x \geq -1$), which probably dominates  
the bulk of matter in the observed Universe. Examples 
of these models include the evolving scalar field 
(Ratra \& Peebles 1988; Frieman et al. 1998; 
Caldwell et al. 1998), the smooth  noninteracting component 
(xCDM) (Turner \& White 1997; Chiba et al. 1997), and still the 
frustated network of topological defects in which $\omega_x = - \
\frac{n}{3}$, being $n$ the dimension of the defect  
(Spergel \& Pen 1997). Some observational aspects of these models 
have extensively been analyzed in the literature. 
For example, Waga \& Miceli (1999), combining statistics of 
gravitational lenses and SNe Ia data have found $\omega_x < -0.7$ 
($68\%$ cl) for a spatially flat Universe. Efstathiou (1999), by using high-z Type Ia 
supernovae and cosmic microwave background anisotropies, 
has found $\omega_x < -0.6$ (2$\sigma$) if the Universe is assumed 
to be spatially flat, or $\omega_x < -0.4$ (2$\sigma$) for 
universes of arbitrary spatial curvature. Perlmutter et al. (1999)  constrained 
$\omega_x < -0.6$ ($95\%$ cl) using large-scale structure and SNe Ia in a
spatially flat geometry.  However, although carefully investigated in many of
their theoretical and  observational aspects, the influence of a 
``quintessence"  component in some kinematic tests like the angular 
size-redshft relation still remains to be analyzed. In principle, the lensing effect of the 
expanding Universe may provide strong limits on the free parameter
describing this exotic component. Therefore, it is interesting to explore how
uncertaints in distance measures of  extragalactic objects and their
underlying  evolutionary effects may  alter the standard cold dark matter
results.  

On the other hand, the existing angular size data for distant objects  are until nowadays 
somewhat controversial, specially because they envolve at least two kinds of
observational dificulties. First, any high redshift object may have a wide
range of proper sizes, and, second, evolutionary and selection effects
probably are not negligible. Indeed, the $\Theta(z)$  relation for some
extended sources samples seems to be quite imcompatible with the predictions
of the  standard FRW model when the latter effects are not taken into account
(Kapahi 1987;1989). There have also been some claims
that the best fit model for the observed distribution of high redshifts
extended objects is provided by the standard Einstein-de Sitter  universe
($q_o={1 \over 2}$, $\Omega_\Lambda=0$) with no significant  evolution
(Buchalter et al. 1998). However, all these results are in contradiction with
the recent observations from type Ia supernovae. Indeed, such data  seem to
ruled out world models filled only by baryonic matter, and more generally, any
model with positive deceleration parameter. The same happens with the
corresponding bounds using the ages of old high redshift galaxies (Dunlop et
al. 1996; Krauss 1997; Alcaniz \& Lima 1999).

The case for compact radio sources is also of great interest.
These objects seem to be less sensitive to evolutionary effects since
they are short-lived ($\sim 10^{3} yr$) and much smaller than their host
galaxy. Initially, the data from a sample of 82  objects gave  remarkable suport for the 
Einstein-de Sitter Universe (Kellerman 1993). However,
some analysis suggest that, although compatible with an 
Einstein-de Sitter Universe, the Kellerman data cannot rule out 
a significant part of the $\Omega_{M}-\Omega_{\Lambda}$ plane 
(Kayser 1995). Some
authors have also argued that models where $\Theta(z)$
diminishes and after a given $z$ remains constant may also
provide a good fit to Kellerman's data. In particular, by
analysing a subset of 59 compact sources within the same
sample,
Dabrowski et al. (1995) found that no useful bounds on the
value of the
deceleration parameter $q_o$ can be derived. Indeed, even
considering that
Euclidean angular sizes ($\Theta \sim z^{-1}$) are excluded at
99$\%$
confidence level, and that the data are consistent with
$q_o=1/2$, they
apparently do not rule out higher values of the deceleration
parameter (Stephanas \& Saha 1995). More recently, based
in a more
complete sample of data, which include the ones originally
obtained by
Kellermann, it was argued that the $\Theta(z)$ relation may be
consistent with any model of the FRW class with deceleration
parameter $\leq 0.5$ (Gurvits et al. 1999).

In this context, we discuss the influence of a ``quintessence" 
component (Q-model) on the angular size-redshift relation. Particular emphasis is 
given for the critical redshift at which the angular 
size of an extragalactic source takes its minimal value. In the limiting case 
($\omega_x=-1$), the results previously derived by Krauss \& Schramm (1993)
for a flat universe with cosmological constant ($\Lambda$CDM) are recovered.
For comparison, we also consider the case of an open model dominated by
nonrelativistic matter (OM).

\section{Angular size and ``quintessence''}

Let us now consider the FRW line element $(c=1)$
\begin{equation}
 ds^2 = dt^2 - R^{2}(t) [d\chi^{2} + S^{2}_{k}(\chi) (d
 \theta^2 +
\rm{sin}^{2} \theta d \phi^{2})]   \quad  ,
\end{equation}
where $\chi$, $\theta$, and $\phi$ are dimensionless comoving
coordinates,  $R(t)$ is the scale factor, and $S_{k}(\chi)$
depends on
the curvature parameter ($k=0$, $\pm 1$). The later function is
defined
by one of the following forms: $S_k (\chi) = \rm{sinh} (\chi)$,
$\chi$,
$\rm{sin} \chi$, respectively, for open, flat  and closed
Universes.

In this background, the angular size-redshift relation for a
rod of intrinsic length $D$ is easily obtained by integrating
the spatial part of the above expression for $\chi$ and $\phi$
fixed. One finds
\begin{equation}
\theta(z) = {D (1 + z) \over R_{o}S_{k}(\chi)}  \quad  .
\end{equation}
The dimensionless coordinate $\chi$ is given by
\begin{equation}
\chi(z) = {1 \over H_o R_o} \int_{(1 + z)^{-1}}^{1} {dx \over x
E(x)}
 \quad  ,
\end{equation}
where $x = {R(t) \over R_o} = (1 + z)^{-1}$ is a convenient
integration variable.
For flat Q-models, the
dimensionless function $E(x)$ takes the following form

\begin{equation}
E_{Q}(x) = \left[(1 - \Omega_{x})x^{-1} +
\Omega_{x}x^{-(1 + 3\omega_x)}\right]^{{1}\over{2}}  \quad  ,
\end{equation}
where $\Omega_{x} =
{8 \pi G\rho_{x} \over 3 H_{o}^{2}}$ is the present day
density parameter associated with the
``quintessence" component. Observe that the flat constraint 
condition, $\Omega_M + \Omega_x =1$, where $\Omega_M = 
{8 \pi G\rho_{M} \over 3 H_{o}^{2}}$ has been explicitly used in the derivation of (4).

Before proceed further, it is interesting to make explicit the connection with some special 
cases already established in the literature. If $\Omega_x = 0$, or still if
$\Omega_x = 1$ and $\omega_x = 0$, one obtains from (2)-(4) the  angular
diameter expression of the Einstein-de Sitter universe (Sandage 1988)
\begin{equation}   
\Theta(z) = {DH_o(1 + z)^{\frac{3}{2}} \over 2\left[(1 +
z)^{1 \over 2} - 1\right]} \quad.   
\end{equation}  
If $\omega_x = -1$, the
Q-model reduces to a $\Lambda$CDM universe, the details of which has been
analysed by Krauss \& Schramm (1993). In particular, if the pair $(\Omega_x,
\omega_x)=(1, -1)$,  the angular diameter of this Q-model is the same of a
flat universe with a pure cosmological constant, namely    
\begin{equation} 
\Theta(z) = {DH_o(1 + z) \over z} \quad.   
\end{equation}
We recall that expression (5) yields a well-known result that the angular diameter in 
Einstein-de Sitter model has a minimum at $z_m=5/4$ (Hoyle 1959), whereas (6)
shows us that the extreme Q-model, ($\Omega_x,\omega_x$)=($1, -1$), has no
minimum at all ($z_m=\infty$). Indeed, for any expanding FRW type cosmology,
the typical behavior of the angular size relation is the existence of a
critical redshift greater than the above  Einstein-de Sitter value.  We also
observe that a new analytical result is obtained by taking $\omega_x=-1/3$ for
 arbitraries  values of $\Omega_x$. From (2)-(4) one finds

\begin{eqnarray} 
\Theta(z) & = &{DH_o(1 + z) \over 2 \sqrt{\Omega_{x}}} \{\rm{ln}[\sqrt{\alpha} + 
\sqrt{\sqrt{\alpha} + 1} \nonumber \\
&    &- {\sqrt{\alpha} \over \sqrt{(1 + z)}} - \sqrt{{\sqrt{\alpha}
\over \sqrt{(1 + z)} }+ 1} ]  \} ^{-1}.
\end{eqnarray}
where $\alpha = {\Omega_x \over { 1 - \Omega_x}}$.  
 
The expression (2) for $\Theta(z)$ cannot be written in simple analytical form, unless the pair 
of parameters ($\Omega_x$, $\omega_x$) take the above mentioned values. For
generic cases, the results can be obtained only by numerical treatment. 

\begin{figure}   
\vspace{.2in}  
\centerline{\psfig{figure=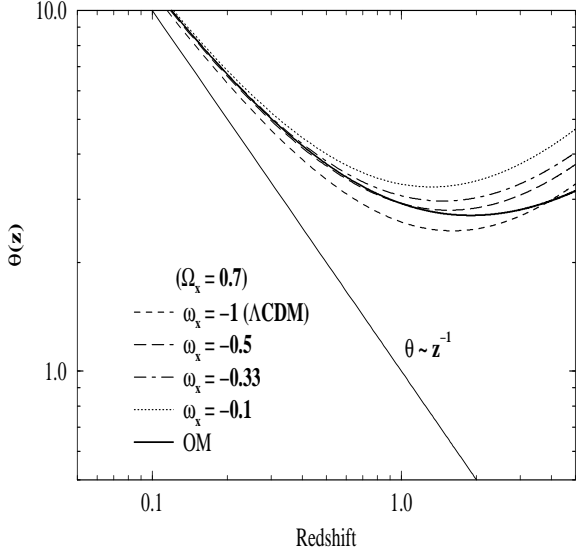,width=3truein,height=3truein}   
\hskip 0.1in}   
\caption{Angular diameter for flat Q-models. Thick solid  curve is the
prediction for an open universe ($\Omega_M = 0.3$) with null cosmological
constant.}      
\end{figure}

In Fig.~1 we show a log-log plot of angular size versus redshift for flat Q-models with 
$\Omega_x = 0.7$ and some selected values of $\omega_x$. For comparison we
have also considered the standard OM cosmology ($\Omega_M = 0.3$). As can be
seen there, for all values of $\omega_x$,  the angular size initially
decreases  with increasing $z$, reaches its minimum value at a given  $z_{m}$,
and eventually begins to increase for fainter magnituds. Note also that the
standard OM behavior may be interpreted as an intermediary case between
$\Lambda$CDM ($\omega_x = -1$) and a Q-model with $\omega_x \leq -0.5$, though its 
critical redshift is displaced to higher values.   

\section{The critical redshift}

As widely known, the existence of a critical redshift $z_m$ on the angular size-redshift relation 
may qualitatively be understood in terms of an expanding space. The light
observed today from a source at high $z$ was emitted when the object was
closer. The relevant aspect here is how this effect may be quantified in terms
of the $\omega_x$ parameter. To analyze the sensivity of the critical redshift
to ``quintessence", we addopt here an approach different of the one applied by
Krauss \& Schramm (1993) to the case of a flat $\Lambda$CDM universe. 

The redshift $z_{m}$ at which the angular size takes its
minimal value is the one cancelling out the derivative of $\Theta$ with
respect to $z$. Hence, from (2) we have the condition

\begin{figure} 
\vspace{.2in} 
\centerline{\psfig{figure=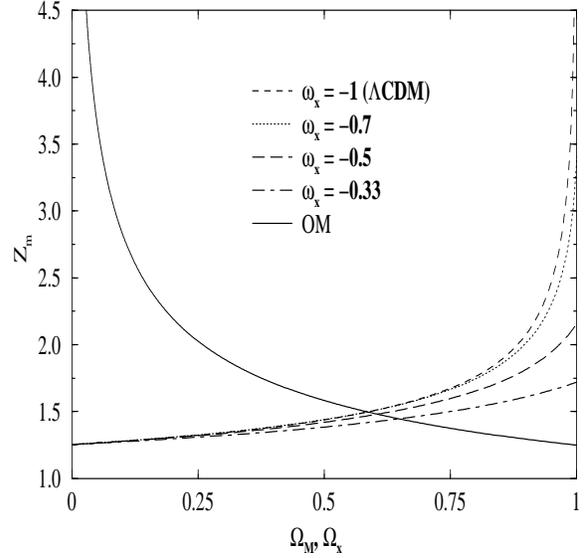,width=3truein,height=3truein} 
\hskip 0.1in} 
\caption{Critical redshift $z_m$ for flat Q-models as a function of $\Omega_x$
and  some selected values of $\omega_x$. The solid line is the prediction of
the open cold dark matter (OM) cosmology.}  
\end{figure}

\vspace{0.5cm}

\begin{equation}
S_k (\chi_m) = (1 + z_m)S'_k (\chi_m)  \quad  ,
\end{equation}
where $S'_k (\chi) = {\partial S_{k} \over \partial
\chi}{\partial
\chi \over \partial z}$, a prime denotes differentiation with
respect
to $z$ and by definition $\chi_{m}= \chi(z_{m})$. 
Observe also that (3) can readily be differentiated yielding
\begin{equation}
(1 + z_{m})\chi'_{m}  =  (R_o H_o)^{-1}  S_{Q}(\Omega_{x},
\omega_{x}, z_m) \quad,
\end{equation}
where
\begin{eqnarray}
S_{Q}(\Omega_{x},
\omega_{x}, z_m) & = & [(1 - 
\Omega_{x})(1 + z_m) + \nonumber \\
    &    & \Omega_{x}(1 + z_m)^{1 + 3
\omega_x}]^{{1\over 2}} \quad. 
\end{eqnarray}

Now, combining equations (8)-(10), we find
\begin{equation}
\int_{(1 + z_m)^{-1}}^{1} {dx \over x E_{Q}(x)} =
S_{Q}(\Omega_{x},
\omega_{x}, z_m)  \quad  .
\end{equation}

The meaning of the above equation is self evident. It
represents an  implicit integro-algebraic equation for the critical
redshift $z_m$ as a function of the parameters defining the flat Q-models. In general, 
this expression cannot be solved in closed
analytical form for $z_m$. However, by taking the limit
$\Omega_x = 0$ in (11), the value $z_m = 1.25$ is readily obtained as
should be expected. The interesting
point here is that (11) is quite convenient for a numerical
treatment. A similar equation can also be derived for an open cold dark matter
universe (OM). We find 
 \begin{equation} 
{\Delta}^{-1}{\rm{tanh}}\left[\Delta \int_{(1 +
z_m)^{-1}}^{1}  {dx \over x
E_{OM}(x)}\right] =
F_{OM}(\Omega_{M}, z_m) ,
\end{equation}
where $\Delta = (1 - \Omega_{M})^{1 \over 2}$ and the functions 
$E_{OM}$, $F_{OM}(\Omega_{M}, z_m)$ are given by
\begin{equation}
E_{OM}(x) = \left[1 - \Omega_{M} + \Omega_{M}
x^{-1}\right]^{{1
\over 2}} 
\end{equation}
\begin{equation}
F_{OM}(\Omega_{M}, z_m) = \left[1 -
\Omega_M  + 
\Omega_M (1 + z_m)\right]^{{1 \over 2}} .
\end{equation}

In Fig.~2 we show the diagrams of $z_m$ as a function of the
density parameter $\Omega_x$, and some selected values of $\omega_x$ (at this point the 
reader should compare our results with the alternative numerical method
developed by Krauss \& Schramm (1993) for a $\Lambda$CDM universe). Note that
equation (12) has also been used to plot the case for the open universes
(solid line). In the former case, the curves show us clearly that all the
Q-models belongs to the same class, which contains the case of a pure
cosmological constant. The smallest value of the critical redshift is exactly
the one given by Einstein-de Sitter universe ($\Omega_x=0$). This value is
pushed to the right direction, that is, for any value of $\omega_x$ it is
displaced to higher redshifts as the $\Omega_x$ parameter increases. For
instance, consider that $\omega_x=-0.5$.  By taking $\Omega_x=0.5$ and
$\Omega_x=0.8$, we find $z_m=1.42$ and $z_m = 1.65$, respectively. In the
opposite extreme ($\Omega_x \rightarrow 1$) the critical redshift is finite
unless the parameter of the equation of state take the extreme value for a
pure $\Lambda$ ($\omega_x=-1$). Note also that for a given  value of
$\Omega_{x}$, the minimum is also displaced for higher redshifts when the 
$\omega_{x}$ parameter diminishes.  However, we see that the effect is small
if the density parameter $\Omega_x$ is low, say, smaller than 0.2, since the
curves are nearly flat and practically coincid below this limit. This
coincidence is even more surprizing for $\omega_x \leq -0.7$. Note that open
models driven by cold dark matter (OM) affects strongly the angular size,
however, in a somewhat different manner as compared to  what happens in a
generic Q-model (see Fig.2). In particular, if the density parameters are
small, say, $\Omega_M$, $\Omega_x$, smaller than 1/3, the critical redshift is
much bigger in the former than in the later. This is easy to understand
physically, because at this limit the contibution of the quintessence is
small, leading to results close to the flat Einstein-de Sitter universe. 

For a large class of Q-models considered in this letter, the critical 
redshifts $z_m$ are displayed 
in Table 1. The OM results have also been quoted for comparison.
The third column with $\omega_x = -1$ corresponds to a flat $\Lambda$CDM model 
(see Krauss \& Schramm 1993). From these results one arrives to a inevitable
conclusion: even neglecting  evolution, the redshift at which the angular
size is minimal cannot alone discriminate between world models since different
scenarios may provide the same $z_m$ values. In particular, for the
observationally favoured open universe ($\Omega_M=0.3$) we find $z_m=1.89$,  a
value that may also be obtained for Q-models having  $0.85 \leq \Omega_x \leq
0.93$ and $-1 \leq \omega_x \leq -0.5$. However, if the angular diameter
results are combinated with other tests, some interesting
cosmological constraints may be obtained. For instance, at galactic scales,
the observed COBE normalized pattern of density fluctuations is more difficult to 
fit within a low-density open universe than in Q-models (Caldwell et al.
1998). Another real possibility is that the universe is actually in an
accelerated expansion state ($q_{o} < 0$), as indicated recently by
measurements using Type Ia  supernovae (Riess et al. 1998; Perlmutter et al.
1998). In this case, any model of the standard FRW class is ruled out
regardless of its curvature parameter. However, the bidimensional parameter
space ($\Omega_x, \omega_x$) is still large enough to accomodate Q-models
predicting both an acelerated expansion (if $q_o < -{1 \over 3\Omega_x}$), and
high values of the critical redshift, say, close to the values of $z_m$ given
by the open models.

\begin{table}[t]  
\begin{center}  
\begin{tabular}{rrrlll}  
\hline  \hline \\
\multicolumn{1}{c}{$\Omega_{M}$}&
\multicolumn{1}{c}{$\frac{\rm{OM}}{(z_{m})}$}& 
\multicolumn{1}{c}{$\frac{\omega_x = -1}{(z_{m})}$}& 
\multicolumn{1}{c}{$\frac{\omega_x = -0.7}{(z_{m})}$}& 
\multicolumn{1}{c}{$\frac{\omega_x = -0.5}{(z_{m})}$}& 
\multicolumn{1}{c}{$\frac{\omega_x = -0.33}{(z_{m})}$}\\  \\  \hline  \hline
1.0& 1.25& 1.25& 1.25& 1.25& 1.25\\ 
0.9& 1.29& 1.28& 1.28& 1.27& 1.27\\
0.8& 1.34& 1.31& 1.31& 1.30& 1.29\\ 
0.7& 1.41& 1.34& 1.34& 1.33&1.32\\  
 0.6& 1.48& 1.38& 1.38& 1.37& 1.35\\ 
0.5& 1.58& 1.44& 1.44& 1.42& 1.38\\  
0.4& 1.71& 1.50& 1.50& 1.47& 1.42\\
0.3& 1.89& 1.60& 1.60& 1.55& 1.47\\ 
0.2& 2.20& 1.76& 1.74& 1.65& 1.53\\
0.1& 2.82& 2.08& 2.0& 1.81& 1.61\\
0.0& $\infty$& $\infty$& 3.27& 2.16& 1.72\\
\hline  \hline
\end{tabular} 
\caption{Critical redshift $z_m$ for some selected 
values of the parameter $\omega_x$. We also quote the results for an open cold dark 
matter model (OM). For $\omega_x=-1$, the results
of the flat Q-models are the same of a  flat $\Lambda$CDM universe previously
derived by Krauss \& Schramm (1993) using a different method.}  
\end{center} 
\end{table} 

The same analytical procedure developed here may be applied when evolutionary and/or 
selection effects due to a linear size-redshift or to a
luminosity-redshift dependence are taken into account{\footnote{For a more
detailed discussion on these effects see Buchalter et al. 1998}} . As widely
believed, a plausible way of standing for such effects is to consider that the
intrinsic linear size has a similar dependence on the redshift as the
coordinate dependence, i.e., $D = D_o (1 + z)^{c}$, being $c < 0$ (Ubachukwu
1995; Buchalter et al. 1998).  In this case, Eq.(11) is still valid but the
function $S_{Q}(\Omega_{x}, \omega_{x}, z_m)$ must be divided by a factor $(1
+ c)$. The displacement of $z_{m}$ relative to the case with  no evolution ($c
= 0$) due to the effects above mentioned may be unexpectedly large.  For
example, if one takes $c = -0.8$ as found by Buchalter et al. 1998, the
redshift of the minimum angular size for the Einstein-de Sitter case 
($\Omega_{x} = 0$) moves from $z_{m} = 1.25$ to $z_{m} = 11.25$. In this way,
the minimal is clearly removed for all practical purposes. This result may be
a possible explanation  why the data of Gurvits et al. (1999), although
apparently in agreement with the Einstein-de Sitter universe, do not show
clear evidence for a minimal angular size close to $z = 1.25$, as should be
expected for this model. This  sort of effect is even greater when an
additional ``quintessence" component is also considered. In the same
vein, since evolution is not forbidden from any principle, we stress that 
constraints from angular size redshift relation should be taken with some
caution.

\vspace{0.3cm} 

{\bf Acknowledgments:} This work
was partially supported by the project Pronex/FINEP (No. 41.96.0908.00) and
Conselho Nacional de Desenvolvimento Cient\'{\i}fico e Tecnol\'ogico - CNPq
(Brazilian Research Agency).

\end{document}